\documentclass[aps,prb,twocolumn,showpacs,superscriptaddress,groupedaddress]{revtex4}  
\usepackage{graphicx}  
\usepackage{dcolumn}   
\usepackage{bm}        
\usepackage{amssymb}   
\usepackage{amsmath}
\usepackage{epstopdf}
\usepackage{ragged2e}

\newcommand {\be}{\begin{eqnarray}}
\newcommand {\ee}{\end{eqnarray}}  
\newcommand {\ba}{\be \begin{aligned}} 
\newcommand {\ea}{\end{aligned} \ee}
\newcommand {\bx}{\mathbf{x}}
\newcommand {\bk}{\mathbf{k}}
\newcommand {\bq}{\mathbf{q}}

\begin{document}

\title{Incommensurate spin density wave at a ferromagnetic quantum critical point in a three-dimensional parabolic semimetal}
\author{James M. Murray}
\affiliation{National High Magnetic Field Lab and Department of Physics, \\
	Florida State University, Tallahassee 32310, USA}
\author{Oskar Vafek}
\affiliation{National High Magnetic Field Lab and Department of Physics, \\
	Florida State University, Tallahassee 32310, USA}
\author{Leon Balents}
\affiliation{Kavli Institute for Theoretical Physics, University of California, 
	Santa Barbara, CA 93106, USA}
\date{\today}

\begin{abstract}
\noindent We explore the ferromagnetic quantum critical point in a three-dimensional semimetallic system with upward- and downward-dispersing bands touching at the Fermi level. Evaluating the static spin susceptibility to leading order in the coupling between the fermions and the fluctuating ferromagnetic order parameter, we find that the ferromagnetic quantum critical point is masked by an incommensurate, longitudinal spin density wave phase. 
We first analyze an idealized model which, despite having strong spin-orbit coupling, still possesses $O(3)$ rotational symmetry generated by the total angular momentum operator. In this case, the direction of the incommensurate spin density wave propagation can point anywhere, while the magnetic moment is aligned along the direction of propagation.
Including symmetry-allowed anisotropies in the fermion dispersion and the coupling to the order parameter field, however, the ordering wavevector instead breaks a discrete symmetry and aligns along either the [111] or [100] direction, depending on the signs and magnitudes of these two types of anisotropy.
\end{abstract}

\maketitle

For decades it has been understood that electron interactions may lead to interesting consequences at low energy scales in three-dimensional (3D) semimetallic systems, in which spin-orbit coupling combines with cubic crystalline symmetries, leading to a band degeneracy at the Fermi energy.\cite{luttinger56,sherrington68,abrikosov71,abrikosov74,moon13,herbut14,savary14,janssen15} Such a band structure has long been known to occur in HgTe and $\alpha$-Sn (gray tin), and has possibly been discovered more recently in the pyrochlore Pr$_2$Ir$_2$O$_7$.\cite{machida09,nakatsuji06,balicas11,kondoXX,rhim15,witczak-krempa14} Since the low-energy effective theory for these systems was first developed by Luttinger\cite{luttinger56}, it has been argued using various approaches that interactions could lead to an excitonic instability at low temperatures\cite{sherrington68,herbut14}, or that the system may be described by an exotic non-Fermi liquid phase, characterized by nontrivial power-law scaling of various physical quantities with temperature.\cite{abrikosov71,abrikosov74,moon13}
In more recent work, the properties of such a system near an Ising antiferromagnetic quantum critical point were explored, and the critical theory was found to be governed by unusual critical exponents and emergent spatial anisotropy of the fermion dispersion.\cite{savary14} The theory describing the quantum phase transition into an insulating nematic phase has also been developed recently.\cite{janssen15}

In this work we explore the fate of a 3D semimetal in the vicinity of a ferromagnetic (FM) quantum critical point. Working at zero temperature and approaching the quantum critical point from the paramagnetic side, we find that it is unstable toward an incommensurate spin density wave (SDW) phase. In the most symmetric version of the theory, there is a combined $O(3)$ rotational symmetry for spin and spatial degrees of freedom, which is spontaneously broken by the SDW wavevector. The $O(3)$ symmetry is reduced to a discrete symmetry by either of two sources of anisotropy: the Yukawa coupling of the fermions to the fluctuating FM field, or the dispersion of the fermions themselves.(Throughout this work, by {\em anisotropy} we shall mean terms preserving the lattice point group symmetries, but not $O(3)$ symmetry.) Either of these sources of anisotropy reduces the $O(3)$ symmetry breaking to a discrete symmetry breaking, with the ordering wavevector lying along one of several high-symmetry directions. In contrast to the more typical case of an incommensurate SDW driven by Fermi surface nesting, the incommensurability of the SDW found in our work is unrelated to Fermi surface or doping effects. Rather, the ordering wavevector depends on nonuniversal parameters such as the temperature and strength of the boson-fermion coupling.

In the following section we introduce the model Hamiltonian for a 3D parabolic semimetal coupled to a fluctuating FM order parameter field and discuss its relevant symmetries. In Section \ref{sec:PiFM} we use this model to study the effects of fermionic fluctuations on the bosonic order parameter field, beginning with the $O(3)$-symmetric case and then moving on to the anisotropic case, with details of these calculations provided in the Appendix. In both cases we obtain a negative contribution to the boson self-energy that is linear in momentum, leading to an instability to an incommensurate SDW phase. In Section \ref{sec:free_energy} we investigate the possibility of a fluctuation-induced first-order phase transition directly from the paramagnetic to the FM phase, evading both the FM quantum critical point and the incommensurate SDW phase. Computing the free energy of the FM order parameter, we find that the sign of the fluctuation-induced nonanalytic term is such that no first-order transition results, in contrast with the metallic case in which there is a nonvanishing density of fermions at the Fermi energy.\cite{belitz97,maslov06,efremov08} Finally, in Section \ref{sec:discussion} we discuss the results and their possible relevance to real materials.


\section{The model}
\begin{figure}
\centering
\includegraphics[width=0.45\textwidth]{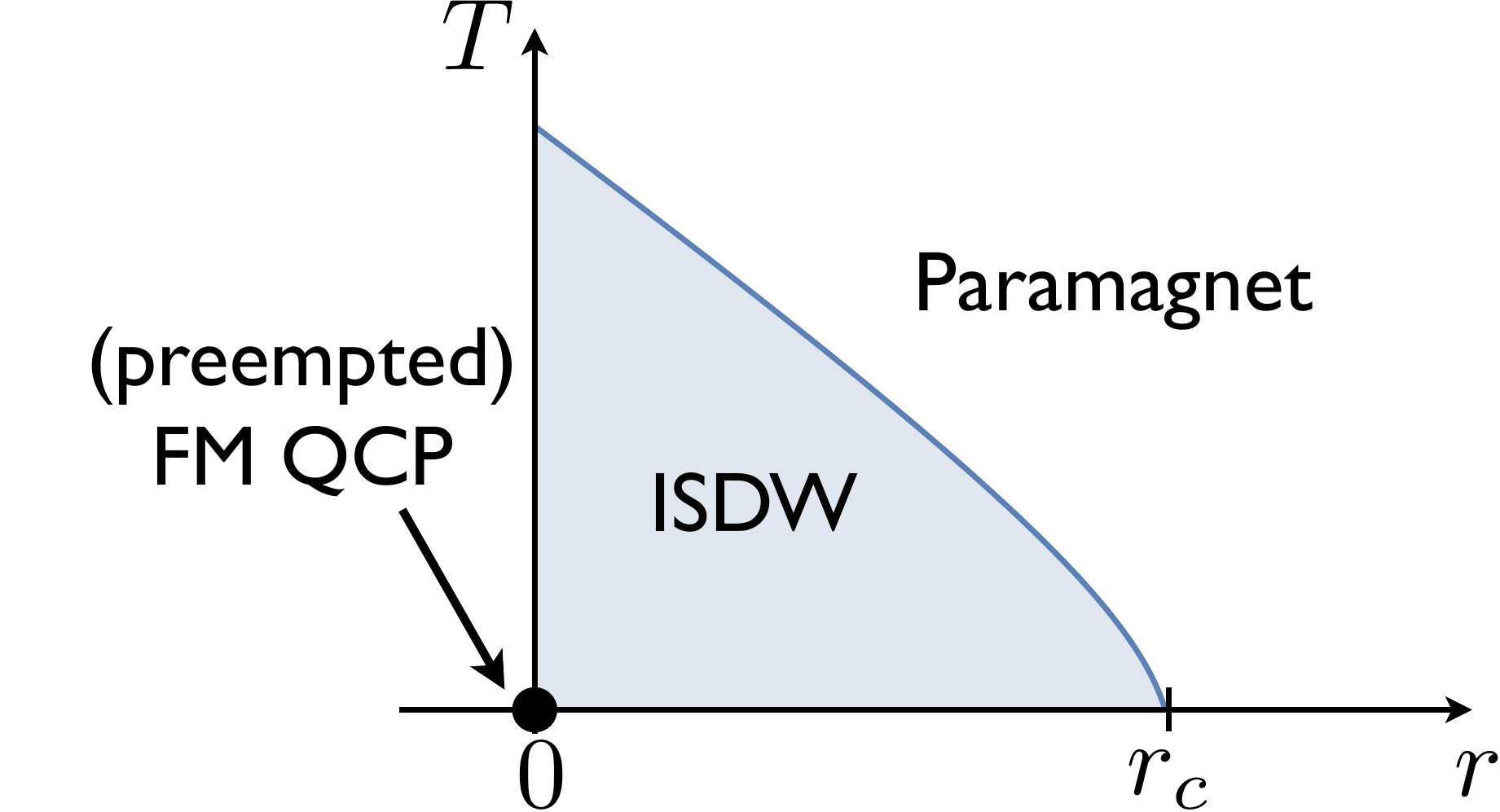}
\caption{Schematic phase diagram showing the onset of incommensurate spin density wave (ISDW) order that preempts the quantum phase transition to a ferromagnetic (FM) phase at $r<0$. The nature of the phase transition between the ISDW and FM phases (not shown) is not specified by our theory.
\label{fig:0514a}}
\end{figure}

The following action describes a theory of fermions coupled to a fluctuating FM order parameter $\vec{\phi}$ at temperature $T=0$:
\ba
\label{eq:0928a}
\mathcal{S} = \int d\tau \; d^3x \bigg\{ &\psi^\dagger [ \partial_\tau + \mathcal{H}_0 (-i\nabla) ] \psi
	+ \vec{\phi} \cdot [\hat G_\phi^{(0)}]^{-1} \cdot \vec{\phi}  \\
&+ \frac{u}{\sqrt{N}} 
	\psi^\dagger ( \vec{M}_1 \cos\alpha +  \vec{M}_2 \sin\alpha) \psi
	\cdot \vec{\phi} ] \bigg\}.
\ea
The fermion fields $\psi$ have $4N$ components, where $N=1$ in the physical case.
The action \eqref{eq:0928a} is similar to the one studied in the Ising antiferromagnetic case\cite{savary14}, but with some important differences. Most obviously, the FM order parameter $\vec{\phi}$ is a vector, rather than a scalar as in the AF case. The mass $r_0 \equiv [G_\phi^{(0)}(0,0)]^{-1}_{ii}$ of this field is tuned to zero at the FM quantum critical point, as shown in Figure \ref{fig:0514a}.
Finally, as pointed out previously\cite{luttinger56,moon13}, there are two symmetry-allowed terms through which FM order can couple to the fermions. Hence $u$ describes the overall strength of the coupling between the fermions and the field $\vec{\phi}$, while $\alpha$ parametrizes the relative strength of the two allowed couplings. 
The $4\times 4$ matrices appearing in \eqref{eq:0928a} are related to the $S=3/2$ spin matrices\cite{luttinger56,murakami04} by $\vec{M}_1 = (S_x, S_y, S_z)$ and $\vec{M}_2 = (S_x^3, S_y^3, S_z^3)$. 

The noninteracting part of the fermion Hamiltonian appearing in \eqref{eq:0928a} is the same as that appearing in Refs.~\onlinecite{murakami04,moon13}:
\be
\label{eq:0928b}
\mathcal{H}_0 (\bk) = c_0 \bk^2 + c_1 \sum_{n=1}^3 d_n(\bk)\Gamma_n + c_2 \sum_{n=4}^5 d_n(\bk)\Gamma_n
\ee
where
\ba
\label{eq:0928c}
d_1(\bk) = \sqrt{3} k_y k_z , \;
d_2(\bk) = \sqrt{3} k_x k_z, \;
d_3(\bk) = \sqrt{3} k_x k_y , \\
d_4(\bk) = \frac{\sqrt{3}}{2}(k_x^2 - k_y^2) , \;
d_5(\bk) = \frac{1}{2} (2k_z^2  -  k_x^2 - k_y^2 ).
\ea
The five $\Gamma$ matrices satisfy $\{ \Gamma_i , \Gamma_j \} = 2 \delta_{ij}$, and one can build ten additional matrices using $\Gamma_{ij} = \frac{1}{2i}[\Gamma_i , \Gamma_j]$. Together with the unit matrix these make a complete basis of $4\times4$ matrices. We use the $\Gamma$-matrix representation, spin matrices $S_i$, and basis functions $d_i (\bk)$ given in Ref.~\onlinecite{murakami04}. For simplicity we assume particle-hole symmetry ($c_0 = 0$), which was found to emerge in the low-energy theory from the renormalization group calculations of Refs.~\onlinecite{moon13,savary14}.

The bare fermion Green function is
\be
\label{eq:0928d}
\hat{G}_0 (i\Omega, \bk) = \frac{-i\Omega 1_4 - \mathcal{H}_0(\bk)}{(i\Omega)^2 - \varepsilon_\bk^2},
\ee
where
\be
\label{eq:0928e}
\varepsilon_\bk = \sqrt{c_1^2 [d_1^2(\bk) + d_2^2(\bk) + d_3^2(\bk)] 
	+ c_2^2 [d_4^2(\bk) + d_5^2(\bk)]}.
\ee
The two-fold degenerate energy bands thus have dispersion $\pm \varepsilon_\bk$, with a quadratic band touching point at $\bk=0$. In the isotropic case in which $c_1 = c_2$, it is straightforward to show that $[\mathbf{L} + \mathbf{S}, \mathcal{H}_0(\bk)] = 0$, where $\mathbf{L} = \mathbf{r} \times \bk$, $[r_i, k_j] = i\delta_{ij}$, and $\mathbf{L} + \mathbf{S}$ is the generator of rotations. In this case the fermionic theory has complete rotational invariance, and the dispersion from \eqref{eq:0928e} becomes simply $\varepsilon_\bk = c_1 \bk^2$.

\section{Ferromagnetic polarization tensor and SDW instability}
\label{sec:PiFM}
In this section we calculate the bosonic polarization tensor, which describes the self-energy of the bosonic field due to damping of spin fluctuations by fermionic excitations. We begin with the isotropic case and then consider the anisotropic case in the following subsections.

\subsection{Polarization tensor: isotropic case}
\label{sec:PiFM_isotropic}
\begin{figure}
\centering
\includegraphics[width=0.45\textwidth]{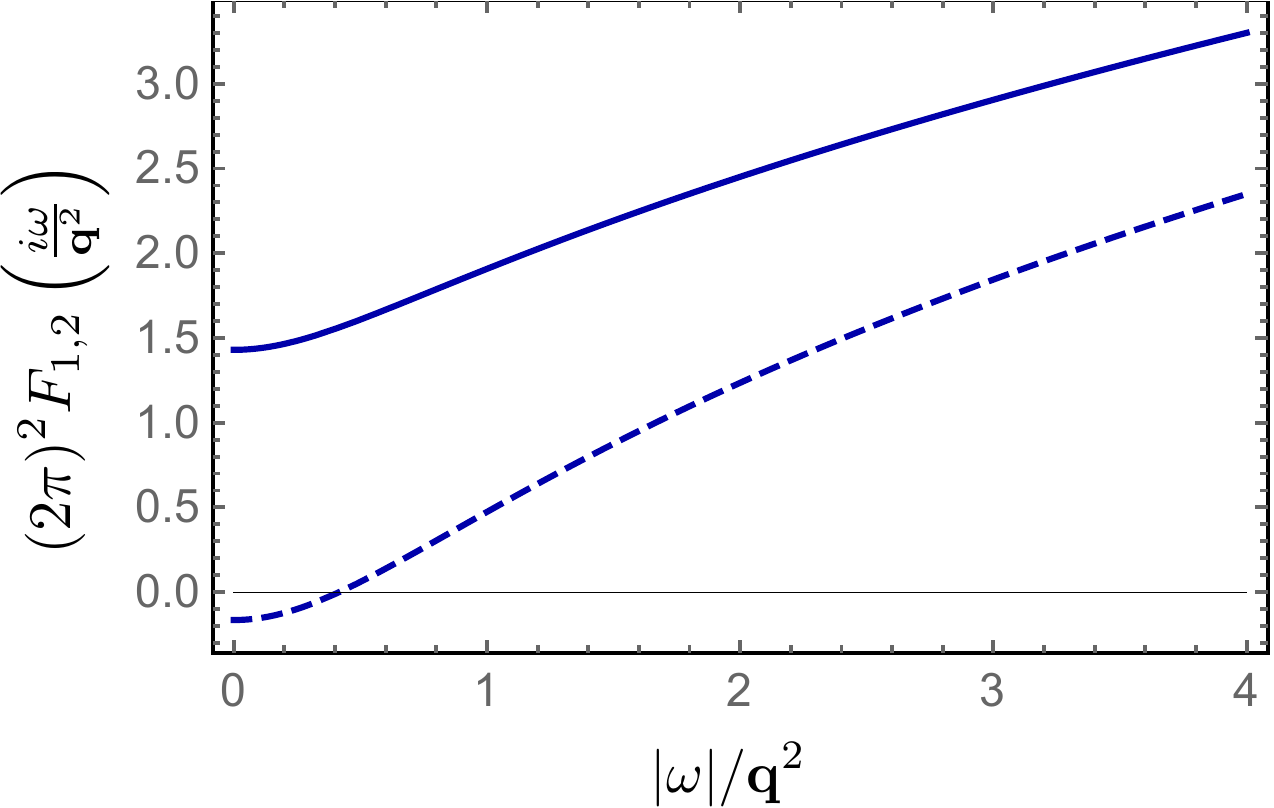}
\caption{The scaling functions appearing in the dressed propagator for the FM field, given by \eqref{eq:0219b}. The solid line shows $F_1 \left( i \Omega / \bq^2 \right)$, and indicates a finite energy cost for transverse spin fluctuations at all frequencies. The dashed line shows $F_2 \left( i \Omega / \bq^2 \right)$, which describes the energy cost of longitudinal spin fluctuations and becomes negative for $|\Omega|/\bq^2 < 0.30$, indicating a phase instability. 
\label{fig:0220a}}
\end{figure}

Let us evaluate the polarization function, first in the relatively simple case where $c_1 = c_2$. Subtracting off the UV-divergent contribution $\Pi_\phi^{ij} (0,0) = - \delta_{ij}u^2\Lambda_c/(2\pi^2 c_1)$ (where $\Lambda_c$ is the ultraviolet momentum cutoff), which can be absorbed into a redefinition of the bosonic mass term by letting $r \equiv r_0 + \Pi_\phi^{ij} (0,0)$, one obtains the following:
\begin{widetext}
\ba
\label{eq:1119a}
\Pi_\phi^{ij} (i\Omega, \bq) &- \Pi_\phi^{ij} (0, 0) = \frac{u^2}{N} \int^\infty \frac{d^3 k}{(2\pi)^3} 
	\int_{-\infty}^{\infty} \frac{d\omega}{2\pi} \bigg\{
	\mathrm{Tr} [ \hat{G}_0 (i\omega + i\Omega/2, \bk + \bq/2) M^i(\alpha) \hat{G}_0 (i\omega - i\Omega/2, \bk - \bq/2) M^j(\alpha) ] \\
	&\quad\quad\quad\quad\quad\quad\quad\quad\quad\quad\quad\quad
	- \mathrm{Tr} [ \hat{G}_0 (i\omega , \bk ) M^i(\alpha)
	\hat{G}_0 (i\omega , \bk) M^j(\alpha) ] \bigg\} \\
&= \frac{u^2}{N} |\bq| \bigg\{ 
	f_1 \left( \tfrac{i\Omega}{\bq^2} \right) \mathrm{Tr} \left[ M^i (\alpha) M^j (\alpha) \right] + \sum_{m,n=1}^5 \bigg[ f_3 \left( \tfrac{i\Omega}{\bq^2} \right)
	(\hat \bq \cdot \Lambda^m \cdot \hat \bq) (\hat \bq \cdot \Lambda^n \cdot \hat \bq )
	+ f_4 \left( \tfrac{i\Omega}{\bq^2} \right)
	\hat\bq \cdot \Lambda^m \Lambda^n \cdot \hat\bq \\
&\quad\quad\quad\quad\quad\quad\quad\quad\quad\quad\quad\quad\quad\quad\quad\quad
	 + f_5 \left( \tfrac{i\Omega}{\bq^2} \right) 
	\mathrm{Tr}[\Lambda^m \Lambda^n]\bigg] 
	\mathrm{Tr} \left[ \Gamma_m M^i (\alpha) \Gamma_n M^j (\alpha) \right]
	\bigg\},
\ea
\end{widetext}
where we have set $c_1 = c_2 = 1$, and defined $\hat\bq = \bq/|\bq|$, $M^i(\alpha) \equiv (M_1^i \cos\alpha + M_2^i \sin\alpha)$, and $\hat\bq\cdot\Lambda^m\cdot\hat\bq = \sum_{ij} \hat q_i \Lambda^m_{ij} \hat q_j$.
The matrices $\Lambda^m$ (not to be confused with the momentum cutoff $\Lambda_c$) are proportional to the five symmetric Gell-Mann matrices and are given in the Appendix, where we also provide the scaling functions $f_i(i\Omega/\bq^2)$. If it is further assumed that $\alpha = 0$, so that the magnetic coupling has complete $O(3)$ symmetry, then \eqref{eq:1119a} becomes the following (noting that the traces in \eqref{eq:1119a} each give a factor of $N$):
\ba
\label{eq:1119b}
&\Pi_\phi^{ij} (i\Omega, \bq) - \Pi_\phi^{ij} (0, 0) 
	\xrightarrow{(\alpha=0)} \\
&u^2 | \bq | \bigg\{ 
	\left[ 5 f_1 \left( \tfrac{i\Omega}{\bq^2} \right) - f_3 \left( \tfrac{i\Omega}{\bq^2} \right)
	- f_4 \left( \tfrac{i \Omega}{\bq^2} \right)
	+ \frac{15}{2} f_5 \left( \tfrac{i \Omega}{\bq^2} \right)
	\right] \delta_{ij} \\
	&\quad\quad\quad+ \left[6 f_3 \left( \tfrac{i\Omega}{\bq^2} \right) 
	+ \frac{21}{2} f_4 \left( \tfrac{i\Omega}{\bq^2} \right) \right] \hat q_i \hat q_j \bigg\}.
\ea
Including the bare kinetic terms, the dressed bosonic stiffness is thus
\ba
\label{eq:0219b}
&[\hat G_\phi (i\Omega, \bq)]^{-1}_{ij} =
	[\hat G_\phi^{(0)} (i\Omega, \bq)]^{-1}_{ij} \\
&\quad\quad\quad + u^2|\bq| \left[ F_1 \left( \tfrac{i \Omega}{\bq^2} \right)
	(\delta_{ij} - \hat q_i \hat q_j) + F_2 \left( \tfrac{i \Omega}{\bq^2} \right) 
	\hat q_i \hat q_j \right],
\ea
where we have introduced
\ba
F_1 &= 5 f_1 \left( \tfrac{i\Omega}{\bq^2} \right) - f_3 			
	\left( \tfrac{i\Omega}{\bq^2} \right)
	- f_4 \left( \tfrac{i \Omega}{\bq^2} \right)
	+ \tfrac{15}{2} f_5 \left( \tfrac{i \Omega}{\bq^2} \right),  \\
F_2 &= 5 f_1 \left( \tfrac{i\Omega}{\bq^2} \right) + 5 f_3 
	\left( \tfrac{i\Omega}{\bq^2} \right)
	+\tfrac{19}{2} f_4 \left( \tfrac{i \Omega}{\bq^2} \right)
	+ \tfrac{15}{2} f_5 \left( \tfrac{i \Omega}{\bq^2} \right).
\ea
The factor in \eqref{eq:0219b} proportional to $(\delta_{ij} - \hat q_i \hat q_j)$ describes the dynamics of transverse spin fluctuations, while the factor proportional to $\hat q_i \hat q_j$ describes longitudinal fluctuations. Figure \ref{fig:0220a} shows the scaling functions appearing in \eqref{eq:0219b}. 
While the transverse part remains positive for all values of $|\Omega|/\bq^2$, implying that these spin fluctuations have a nonzero energy cost at all frequencies, the longitudinal part becomes negative at frequencies $|\Omega| < 0.42 \bq^2$, implying a vanishing energy cost for longitudinal spin fluctuations at sufficiently small frequencies and momenta.

To $O(\bq^2)$, the static part of bare boson propagator has the following analytic form:
\be
\label{eq:0220f}
[\hat G_\phi^{(0)} (0, \bq)]^{-1}_{ij} = r \delta_{ij} +
	\bq^2[ v_1^2 (\delta_{ij} - \hat q_i \hat q_j ) 
	+ v_2^2 \hat q_i \hat q_j ] ,
\ee
with stability of the bare theory requiring that $v_1^2 > 0$ and $v_2^2 > 0$ for $r \geq 0$. From \eqref{eq:0219b} it is evident that if one fixes $\Omega=0$ and decreases momentum $|\bq|$ with $r$ sufficiently small, there will be a critical value of $|\bq|$ where the negative correction $u^2 F_2(0) |\bq|$ will overtake the bare term.
The consequences of this become evident upon inverting the expression \eqref{eq:0219b} at $\Omega=0$ to obtain the static spin susceptibility:
\ba
\label{eq:0219d}
\chi_{ij}(0, \bq) =& \frac{1}{r + v_1^2 \bq^2 
	+ F_1(0) u^2 |\bq|} (\delta_{ij} - \hat q_i \hat q_j) \\
&+ \frac{1}{r + v_2^2 \bq^2 
	+ F_2(0) u^2 |\bq|}	\hat q_i \hat q_j,
\ea
where the constants are $F_1(0) \approx 0.0362$, and $F_2(0) \approx - 0.0042$.
Equation \eqref{eq:0219d} makes it clear that the vanishing longitudinal term for sufficiently small $r$ corresponds to an instability toward a spin density wave phase at finite wavevector $|\mathbf{Q}| = |F_2(0)| u^2 / 2 v_2^2$. This incommensurate instability is similar to the ``spiral SDW'' instability that has been noted in previous theories of itinerant fermion systems near a quantum critical point described using the spin fermion model\cite{chubukov04,rech06}, and to the behavior recently observed in the metallic ferromagnet PrPtAl.\cite{abdul-jabbar15} In the isotropic case that we have so far considered, the wavevector $\mathbf{Q}$ of the SDW order could be along any direction, and thus breaks an $O(3)$ symmetry, with the spins aligning along (since the SDW is {\em longitudinal}) the direction of $\mathbf{Q}$. The fact that the spins cannot independently choose a direction along which to align is a direct consequence of the spin-orbit coupling built into this model, and distinguishes this theory from the spin fermion theory of itinerant ferromagnets, for which the $SU(2)$ spin rotation symmetry is independent of spatial rotations. Finally, we note that while our calculations are performed at $T=0$, one can see that at $T>0$ the temperature dependence will enter through the scaling function $F_2$. This leads to temperature dependence of the ordering wavevector, which is an unusual feature in an SDW system.

\subsection{Polarization tensor: Yukawa anisotropy}
Before proceeding to the fully anisotropic case with both $c_1 \neq c_2$ and $\alpha \neq 0$, in this section we investigate the case in which the Yukawa coupling is anisotropic ($\alpha \neq 0$) but the fermion dispersion remains isotropic ($c_1 = c_2$). In this case the summations in \eqref{eq:1119a} may be performed to obtain a polarization tensor of the following form:
\ba
\label{eq:0514a}
& \Pi_\phi^{ij} ( i\Omega, \bq)  - \Pi_\phi^{ij} (0, 0) = \\
& \quad u^2 |\bq| \bigg(
	\left[ h_1 (\tfrac{i\Omega}{\bq^2},\alpha) 
	+ h_2 (\tfrac{i\Omega}{\bq^2},\alpha) \hat q_i^2
	+ h_3 (\tfrac{i\Omega}{\bq^2},\alpha) \hat q_i^4 \right] \delta_{ij} \\
& \quad\quad\quad + \left[ h_4 (\tfrac{i\Omega}{\bq^2},\alpha) 
	+ h_5 (\tfrac{i\Omega}{\bq^2},\alpha) (\hat q_i^2 + \hat q_j^2) \right] 
	\hat q_i \hat q_j
	\bigg) .
\ea
The scaling functions $h_i$ are linear combinations of the $f_i$ appearing in \eqref{eq:1119a}, with coefficients depending on the Yukawa angle $\alpha$. 

In order to determine the nature of the SDW instability in the anisotropic case, we can obtain the eigenvalues and eigenvectors of the polarization tensor for these two cases from the following equation:
\be
\label{eq:0514b}
\frac{1}{u^2|\bq|} \left[ \hat\Pi_\phi (0, \bq) - \hat\Pi_\phi (0,0) \right]
	\psi_\phi (\hat\bq) = \pi_\phi (\hat\bq) \psi_\phi (\hat\bq).
\ee
For $\bq \sim (1,0,0)$, the eigenvectors are aligned with the principal axes. In this case the longitudinal mode, for which $\psi_\phi(\hat\bq) = (1,0,0)$, has eigenvalue $\pi_\phi^{(L)}(\hat\bq) = h_1 + h_2 + h_3 $, while the two degenerate transverse modes have $\pi_\phi^{(T)}(\hat\bq) = h_1 $. For $\bq \sim (1,1,1)$, on the other hand, the longitudinal mode has eigenvalue $\pi_\phi^{(L)}(\hat\bq) = h_1 + \tfrac{1}{3} h_2 + \tfrac{1}{9} h_3 + h_4 + \tfrac{2}{3} h_5$, while the transverse modes have eigenvalues $\pi_\phi^{(T)}(\hat\bq) = h_1 + \tfrac{1}{3} h_2 + \tfrac{1}{9} h_3$. By tracking the minimum eigenvalue as a function of $\hat\bq$, as shown in Figure \ref{fig:0515b}, we find that the ordering occurs along [111] and is longitudinal for all values of the Yukawa parameter in the range $0 \leq \alpha \leq \pi/2$.

\vspace{0.5cm}
\subsection{Polarization tensor: fully anisotropic case}
Considering now the more general case in which both $c_1 \neq c_2$, and $\alpha \neq 0$, in this subsection we show the way in which these two types of anisotropy favor ordering along certain crystalline axes. The polarization tensor for the FM field $\phi$ in this general case is given by
\begin{widetext}
\ba
\label{eq:0928h}
\Pi_\phi^{ij} (i\Omega, \bq) - \Pi_\phi^{ij} (0,0) 
&= \frac{u^2}{N}|\bq| \bigg\{
	\Phi_1 \left( \tfrac{i\Omega}{\bq^2}, \hat\bq \right)
	\mathrm{Tr} \left[ M^i (\alpha) M^j (\alpha) \right] 
	+ \sum_{n=1}^5 \Phi_2^{(n)} \left( \tfrac{i\Omega}{\bq^2} , \hat\bq \right)
	\mathrm{Tr} \left[ M^i (\alpha) \Gamma_n M^j (\alpha) \right] \\
	&\quad
	+ \sum_{m=1}^5 \Phi_3^{(m)} \left( \tfrac{i\Omega}{\bq^2}, \hat\bq \right)
	\mathrm{Tr} \left[ M^i (\alpha) M^j (\alpha) \Gamma_m \right]
	+ \sum_{m,n=1}^5 \Phi_4^{(m,n)} \left( \tfrac{i\Omega}{\bq^2}, \hat\bq \right)
	\mathrm{Tr} \left[ \Gamma_m M^i (\alpha) \Gamma_n M^j (\alpha) \right]
	\bigg\}.
\ea
\end{widetext}
The scaling functions appearing in \eqref{eq:0928h} are given in the Appendix, Equations \eqref{eq:0220g}--\eqref{eq:0220j}.
Introducing the anisotropy parameter $\delta$ for the fermion dispersion by letting $c_{1,2} = 1 \pm \delta$, these scaling functions can then be obtained perturbatively in $\delta$, as shown in the Appendix. 

As pointed out in the two previous subsections, the behavior of spin excitations becomes increasingly dominated by the contribution from $\hat \Pi_\phi (i\Omega, \bq)$ at small momenta and frequencies. Due to the anisotropy, certain ordering wavevectors will be favored when either of these parameters is nonzero. 
\begin{figure}
\centering
\includegraphics[width=0.48\textwidth]{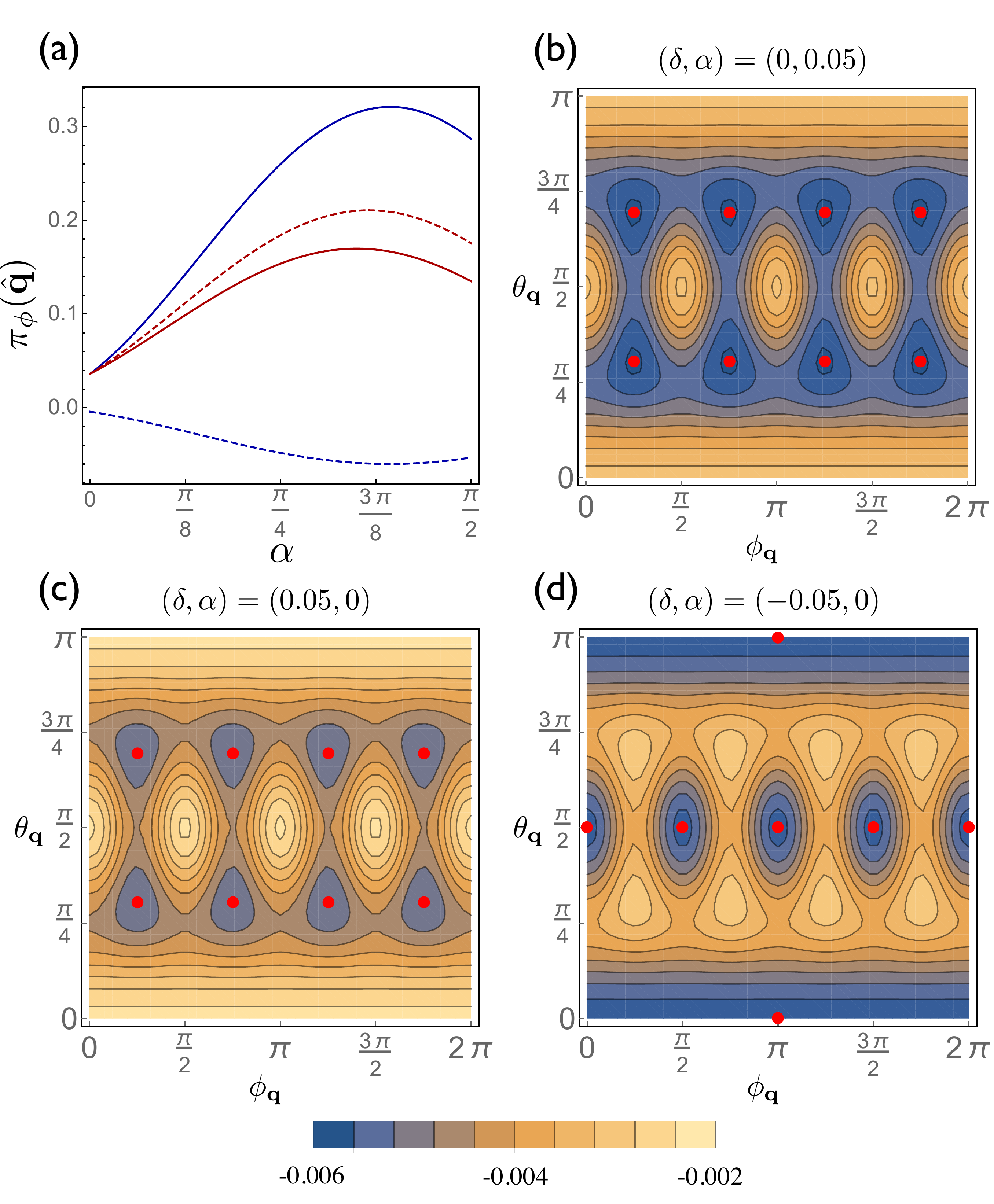}
\caption{(a) The polarization tensor eigenvalues as a function of the Yukawa angle $\alpha$ with $\delta=0$. Blue (red) lines correspond to longitudinal (transverse) modes, while solid (dashed) lines have $\bq \sim [100]$ ([111]). The lowest eigenvalue occurs for the longitudinal mode with $\bq\sim [111]$. (b)-(d) The minimum eigenvalues $\pi_\phi (\hat\bq)$. The direction of the ordering wavevector for the incommensurate SDW will correspond to one of the potential minima, shown as red points. (b) Anisotropy in the Yukawa coupling leads to minima along [111] and equivalent directions. (c) The positive fermion dispersion anisotropy parameter $\delta>0$ leads to minima along [111] and equivalent directions. (d) $\delta<0$ leads to minima along [100] and equivalent directions. 
\label{fig:0515b}}
\end{figure}
In order to determine the wavevector of the SDW instability, we once again investigate the minimum eigenvalue $\pi_\phi(\hat\bq)$ from \eqref{eq:0514b}.
This quantity is plotted in Figure \ref{fig:0515b} as a function of the angle of $\hat\bq$. We find that the ordering occurs along [111] when $\delta > 0$, while $\delta < 0$ favors ordering along [100]. 
With such anisotropy taken into account, the ordered state itself is now an incommensurate SDW with wavevector whose direction is locked to the crystalline axes, and both the zero temperature and $T>0$ transitions will no longer be in the O(3) ferromagnetic class as in the naive unrenormalized theory, but must be reconsidered in this light.  For brevity and to maintain focus we do not do this here.

One can also ask whether the incommensurate SDW instability survives in the case that $\delta$ is not perturbatively small. 
\begin{figure}
\centering
\includegraphics[width=0.45\textwidth]{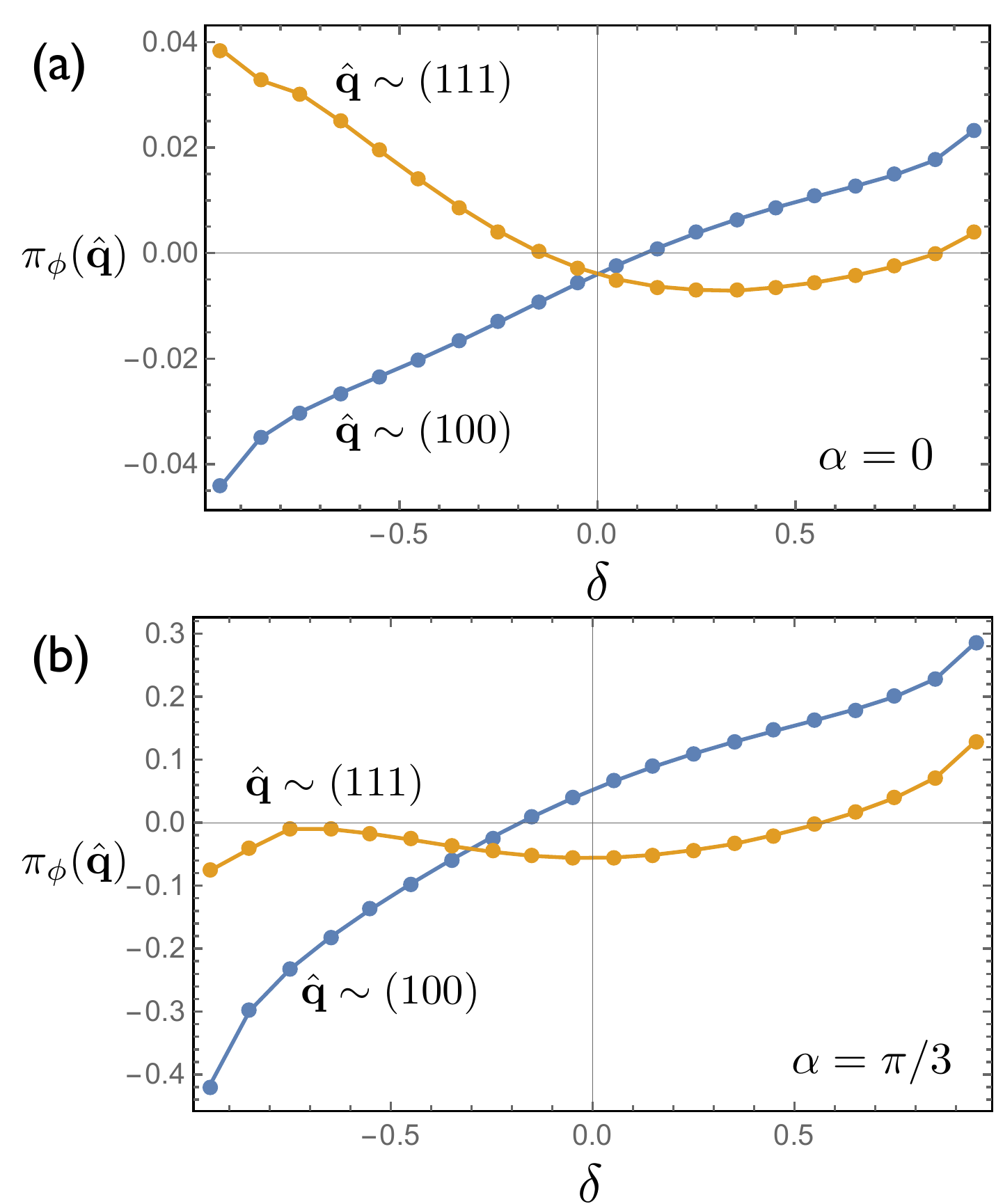}
\caption{Minimum eigenvalue of the polarization tensor as a function of the dispersion anisotropy, with the wavevector $\hat\bq$ along the [100] direction (blue) and [111] direction (orange). The incommensurate SDW instability occurs at the wavevector for which the eigenvalue $\pi_\phi (\hat\bq)$ is most negative. The angle parametrizing the Yukawa interaction is set to $\alpha=0$ and $\alpha = \pi/3$ in (a) and (b), respectively.
\label{fig:0310a}}
\end{figure}
In this case the polarization tensor can be evaluated by directly integrating the scaling functions appearing in \eqref{eq:0928h} numerically at $\Omega = 0$ for a given value of $\delta$. As shown in Figure \ref{fig:0310a}, the instability occurs for most values of $\delta$ when the Yukawa parameter lies in the physically expected range $0 \leq \alpha \leq \pi/2$.
The exception occurs in the case where $\delta$ approaches 1, or equivalently for $c_1 \gg c_2$. 
The physical case, however, is more likely in the opposite limit of $\delta < 0$ (or equivalently $c_1 < c_2$), which is the case appearing most generically for simple tight-binding models on the pyrochlore lattice.\cite{kurita11,witczak-krempa12} Accordingly, the case $c_1 \ll c_2$ was considered in the anisotropic RG theory of Ref.~\onlinecite{savary14}. Also note that we have calculated $\pi_\phi (\hat\bq)$  only for wavevector directions $\hat\bq \sim [100]$ and $\hat\bq \sim [111]$. Based on the results for small $\delta$ shown in Figure \ref{fig:0515b} and general symmetry considerations, it is natural to expect that the ordering will occur along one of these directions, though we have not shown this rigorously.

\section{Free energy expansion}
\label{sec:free_energy}
Previous investigations of itinerant fermionic systems near an FM quantum critical point have found that the quantum phase transition can be precluded by either of two types of instability.\cite{brando15} The first, in which fluctuations contribute a negative term to the bosonic self-energy, leading to an SDW instability at finite wavevector $|\mathbf{Q}|$, was already discussed in the previous section. The second is a fluctuation-induced first-order phase transition into the FM phase, the possibility of which can be inferred by the presence of a nonanalytic term in the order parameter in the free energy expansion.\cite{belitz97,maslov06,efremov08} In this section we derive the Landau free energy for the FM order parameter, showing that, although a fluctuation-induced nonanalytic term is present, its sign is such that no first-order instability results. (The absence of such a transition has also been found in the Ising antiferromagnetic case.\cite{savary14})

The following effective action is obtained by integrating out the fermions from the original action \eqref{eq:0928a}:
\ba
\label{eq:0224a}
S_\mathrm{eff} =& \frac{1}{2} \int d\tau \int d^dx\;
	\left( \vec{\phi}\cdot [\hat G_\phi^{(0)}]^{-1} \cdot \vec{\phi}
	+ b |\vec{\phi}|^4 \right) \\
& - \mathrm{Tr}\; \mathrm{ln} 
	\left[ \hat G_0^{-1} 
	- \frac{u}{\sqrt{N}} \vec{M} (\alpha) \cdot \vec{\phi} \right].
\ea
The trace in the second term is a trace over matrix elements, as well as over frequency and momentum. The quartic term $\sim b |\vec{\phi}|^4$, with $b > 0$, is included since it is allowed by symmetry and arises in the low-energy effective theory from integrating out high-energy modes. Due to the gapless fermionic excitations, one cannot simply follow Hertz\cite{hertz76} and expand the logarithm in powers of $\vec{\phi}$ to obtain an effective theory for the order parameter field. Rather, a useful first step is to differentiate \eqref{eq:0224a} with respect to the order parameter. And because we are interested in uniform states, we can evaluate the result at zero external momentum and frequency, so one has $[\hat G_\phi^{(0)}]^{-1}_{ij} = r \delta_{ij}$, \textit{i.e.}~the bare mass of the bosonic field. The result is
\ba
\label{eq:0224b}
&\frac{\delta F[\vec{\phi}]}{\delta \phi_i} 
	=\frac{\delta S_\mathrm{eff}}{\delta \phi_i}\bigg|_{i\omega = \bk = 0} \\
&= r \phi_i + b \vec{\phi}^2  \phi_i
	+ \frac{u}{\sqrt{N}} \int \frac{d\Omega}{2\pi} \int \frac{d^3 q}{(2\pi)^3} \\
& \quad\quad\quad \times \mathrm{Tr} \bigg[ \bigg( \hat G_0^{-1}(i\Omega, \bq)
	- \frac{u}{\sqrt{N}} \vec{M}(\alpha)\cdot\vec{\phi}\bigg)^{-1} 
	M^i(\alpha) \bigg].
\ea
If we take the matrix inverse and specialize to the isotropic case in which $\alpha = 0$, this becomes
\ba
\label{eq:0224c}
\frac{\delta F[\vec{\phi}]}{\delta \phi_i} 
	&= r \phi_i + b \vec{\phi}^2 \phi_i
	+ \frac{u^2}{N} \phi_i \int \frac{d\Omega}{2\pi} \int \frac{d^3 q}{(2\pi)^3}
	f\left(i\Omega, \bq, \vec{\phi} \right) \\
&= \left[ r  + \frac{u^2}{N} \int_{\Omega, \bq} f(i\Omega, \bq, 0) \right] \phi_i
	+ b \vec{\phi}^2 \phi_i \\
& \quad\quad\quad  + \frac{u^2}{N} \phi_i \int_{\Omega, \bq}
	\left[ f\left(i\Omega, \bq, \vec{\phi} \right)
	- f ( i\Omega, \bq, 0) \right],
\ea
where $f(i\Omega, \bq, \vec{\phi} )$ is a complicated function, and in the second line we have added and subtracted the UV-divergent piece. By rescaling $\Omega \to |\vec{\phi}|\Omega,\; \bq \to \sqrt{|\vec{\phi}|}\bq$ in the second integral, and taking advantage of the fact that---due to rotational symmetry---the integral must only depend on the magnitude of $\vec{\phi}$ but not its direction, \eqref{eq:0224c} becomes
\ba
\label{eq:0224d}
\frac{\delta F[\vec{\phi}]}{\delta \phi_i} 
=& \left[ r +  b \vec{\phi}^2 + \frac{u^2}{N} \int_{\Omega, \bq} f(i\Omega, \bq, 0) \right] \phi_i \\
	& + \frac{u^2}{N} \phi_i \sqrt{|\vec{\phi}|} \int_{\Omega, \bq}
	\left[ f\left(i\Omega, \bq, 1 \right)
	- f ( i\Omega, \bq, 0) \right],
\ea
from which the free energy is
\be
\label{eq:0224e}
F [ \vec{\phi} ] = a |\vec{\phi}|^2 + c |\vec{\phi}|^{5/2} + \tfrac{1}{2} b |\vec{\phi}|^4.
\ee
By evaluating the integral in \eqref{eq:0224d} numerically, the second constant appearing in \eqref{eq:0224e} is $c \approx 0.019 u^2 /N$. Since $c > 0$, we conclude that the nonanalytic term does not lead to a first-order phase transition into the FM phase, and that the FM quantum critical point is preempted by an incommensurate SDW via the mechanism discussed in Section \ref{sec:PiFM}.

\section{Discussion}
\label{sec:discussion}
In this work we have shown that the quantum critical point in a 3D parabolic semimetal near the onset of FM order is preempted by an instability to an incommensurate, longitudinal SDW phase. While the wavevector $\mathbf{Q}$ characterizing the SDW order breaks a continuous symmetry in the idealized, $O(3)$-symmetric version of the theory, including anisotropy terms allowed by crystalline symmetry leads to a discrete symmetry breaking, with $\mathbf{Q}$ along either the [111] or [100] direction, depending on the values of the anisotropy parameters as shown in Figure \ref{fig:0515b}. We have shown that the instability is present over a broad range of anisotropies, apart from the extremely anisotropic case where $c_1 \gg c_2$. Investigating this limit and its RG flows further, along the lines of Ref.~\onlinecite{savary14}, could be an interesting direction for future work. Finally, we have also investigated the possibility of a fluctuation-induced first-order transition into the FM phase, finding that, although a nonanalytic term in the free energy does exist at zero temperature, its sign is such that no first-order instability results. 

While in this work we have not focussed on the role played by Coulomb interaction, some general statements about its effects can be made. As pointed out in previous works\cite{moon13,savary14}, Coulomb interaction tends to decrease the effective mass difference between the electron and hole bands, as well as the rotational anisotropy in the fermion dispersion. Because the SDW instability dominates the low-energy physics, the effects of Coulomb interactions will be secondary near the FM quantum critical point. (A possible exception is in the limit $c_1 \gg c_2$, where---as discussed above---the SDW instability is absent. In this case we cannot rule out the possibility that the combined effects of Coulomb interactions and spin fluctuations could in principle lead to a stable anisotropic fixed point such as the one found in Ref.~\onlinecite{savary14}.)

While there is no obvious existing material described by the critical theory developed in this paper, it is possible that the vicinity of a (preempted) FM quantum critical point could be reached by doping existing parabolic semimetals such as HgTe or $\alpha$-Sn with magnetic impurities such as Mn. Half-Heusler and Heusler materials can exhibit narrow-gap or semimetallic behavior\cite{casper12,chadov10} as well as ferromagnetic critical points\cite{naka12}, although to our knowledge the combination of these two conditions has not been shown to exist in any one material.

\begin{acknowledgements}
OV and JM were supported by the NSF CAREER award
under Grant No.~DMR-0955561, NSF Cooperative
Agreement No.~DMR-1157490 and the State of Florida. LB was supported by the DOE Office of Basic Energy Sciences, Grant DE-FG02-08ER46524.
\end{acknowledgements}
%

\begin{appendix}
\section{}

\begin{widetext}
In this Appendix we derive the scaling functions determining the polarization tensor given in the main text.
In order to compute the FM polarization tensor, we begin by evaluating the following general outer product of fermionic Green functions:
\ba
\label{eq:1128a}
\int \frac{d\omega}{2\pi} \int \frac{d^3 k}{(2\pi)^3} \bigg[ &\hat{G}_0 
	\left( i\omega+i\frac{\Omega}{2}, \bk+ \frac{\bq}{2} \right) 
	\otimes \hat{G}_0 \left( i\omega-i\frac{\Omega}{2}, \bk- \frac{\bq}{2} \right) 
	- \hat{G}_0 \left( i\omega, \bk \right) 
	\otimes \hat{G}_0 \left( i\omega, \bk \right) \bigg] \\
&= |\bq| \bigg[ \Phi_1 (\tfrac{i\Omega}{\bq^2}, \hat\bq) 1_4 \otimes 1_4
	+ \sum_n \Phi_2^{(n)} 
	(\tfrac{i\Omega}{\bq^2}, \hat\bq) 1_4 \otimes \Gamma_n \\
& \quad\quad + \sum_m \Phi_3^{(m)} 
	(\tfrac{i\Omega}{\bq^2}, \hat\bq) \Gamma_m \otimes 1_4
	+ \sum_{m,n} \Phi_4^{(m,n)} (\tfrac{i\Omega}{\bq^2}, \hat\bq) \Gamma_m \otimes \Gamma_n \bigg],
\ea
where the UV-divergent piece has been subtracted off, as described in the main text. Once this quantity is determined, any response function (and, in particular, the response function in the FM channel) can be calculated by contracting \eqref{eq:1128a} with appropriate matrices. The scaling functions appearing in \eqref{eq:1128a} are given by
\be
\label{eq:0220g}
\Phi_1 (\tfrac{i\Omega}{\bq^2}, \hat\bq) = \int\frac{d^3 x}{(2\pi)^3} \left\{ 
	\frac{-(\varepsilon_{\bx+\hat\bq/2} + \varepsilon_{\bx-\hat\bq/2})}
	{2 \left[ (\varepsilon_{\bx-\hat\bq/2} + \varepsilon_{\bx+\hat\bq/2})^2 
	- \left( \tfrac{i\Omega}{\bq^2} \right)^2 \right]}
	+ \frac{1}{4\varepsilon_\bx}
	\right\},
\ee
\be
\label{eq:0220h}
\Phi_2^{(n)} (\tfrac{i\Omega}{\bq^2}, \hat\bq)
	= - \Phi_3^{(n)} (\tfrac{i\Omega}{\bq^2}, \hat\bq) 
	= \hat c_n \int\frac{d^3 x}{(2\pi)^3} \frac{\frac{i\Omega}{\bq^2} 
	d_n (\bx - \tfrac{\hat\bq}{2})}{2 \varepsilon_{\bx-\hat\bq/2} 
	\left[ (\varepsilon_{\bx-\hat\bq/2} + \varepsilon_{\bx+\hat\bq/2})^2 
	- \left( \frac{i\Omega}{\bq^2} \right)^2 \right]} ,
\ee
\be
\label{eq:0220j}
\Phi_4^{(m,n)} (\tfrac{i\Omega}{\bq^2}, \hat\bq) 
	= \hat c_m \hat c_n \int\frac{d^3 x}{(2\pi)^3} \left\{
	\frac{(\varepsilon_{\bx+\hat\bq/2} + \varepsilon_{\bx-\hat\bq/2})
	d_m (\bx + \tfrac{\hat\bq}{2}) d_n (\bx - \tfrac{\hat\bq}{2})}
	{2 \varepsilon_{\bx+\hat\bq/2} \varepsilon_{\bx-\hat\bq/2}
	\left[ (\varepsilon_{\bx-\hat\bq/2} + \varepsilon_{\bx+\hat\bq/2})^2 
	- \left( \frac{i\Omega}{\bq^2} \right)^2 \right]} 
	- \frac{d_m (\bx ) d_n (\bx)}{4 \varepsilon_\bx^3} \right\}.
\ee
From \eqref{eq:0220j} it can be noted that $\Phi_4^{(m,n)} = \Phi_4^{(n,m)}$. In the above equations we have introduced $\hat c_{1,2,3} = c_1$ and $\hat c_{4,5} = c_2$.  

In what follows it is useful to introduce matrices such that
\be
\label{eq:1029g}
d_m\left( \bx \right) = 
	\sum_{i,j=1}^3 x_i \Lambda^m_{ij} x_j,
\ee
where the matrices $\Lambda^m$ are 
\ba
\label{eq:1029f}
\Lambda ^1&=\frac{\sqrt{3}}{2} \left(
\begin{array}{ccc}
 0 & 0 & 0 \\
 0 & 0 & 1 \\
 0 & 1 & 0 \\
\end{array}
\right), \quad
\Lambda ^2=\frac{\sqrt{3}}{2} \left(
\begin{array}{ccc}
 0 & 0 & 1 \\
 0 & 0 & 0 \\
 1 & 0 & 0 \\
\end{array}
\right), \quad
\Lambda ^3=\frac{\sqrt{3}}{2} \left(
\begin{array}{ccc}
 0 & 1 & 0 \\
 1 & 0 & 0 \\
 0 & 0 & 0 \\
\end{array}
\right), \\
\Lambda ^4 &=\frac{\sqrt{3}}{2} \left(
\begin{array}{ccc}
 1 & 0 & 0 \\
 0 & -1 & 0 \\
 0 & 0 & 0 \\
\end{array}
\right), \quad
\Lambda ^5=\frac{1}{2} \left(
\begin{array}{ccc}
 -1 & 0 & 0 \\
 0 & -1 & 0 \\
 0 & 0 & 2 \\
\end{array}
\right).
\ea
These are proportional to the five symmetric Gell-Mann matrices\cite{georgi99} $\lambda_i$:
\be
\frac{2}{\sqrt{3}}(\Lambda^1, \Lambda^2, \Lambda^3, \Lambda^4, \Lambda^5) = (\lambda_6, \lambda_4, \lambda_1, \lambda_3, - \lambda_8).
\ee
We shall also make use of the fact that these matrices are traceless ($\sum_i \Lambda^m_{ii} = 0$).

While it is possible to integrate the scaling functions \eqref{eq:0220g}--\eqref{eq:0220j} numerically for a fixed value of the external momentum $\hat\bq$, more convenient expressions can be obtained by separating out the $\hat\bq$ dependence from these integrals by applying the following transformation:
\ba
\label{eq:1029d}
x_i &= \sum_j [\hat q_i \hat q_j + (\delta_{ij} - \hat q_i \hat q_j) \cos\eta + \epsilon_{ijk} \hat q_k \sin\eta ] x'_j \\
	& \equiv \sum_j A_\eta^{ij} x'_j
\ea
and averaging the integrand over the angle $\eta$. Because this transformation rotates the integration variable $\bx$ around the fixed unit vector $\hat\bq$ by an angle $\eta$, one has $\bx'^2 = \bx^2$ and $\hat\bq\cdot\bx' = \hat\bq\cdot\bx$. The following product of rotation matrices appears when \eqref{eq:1029d} is substituted into the scaling function integrals \eqref{eq:0220g}--\eqref{eq:0220j}:
\ba
\label{eq:0517a}
A_\eta^{ip} A_\eta^{jq} &\to L_{ij} L_{pq} + \frac{1}{2} T_{ij}T_{pq}
	+ (L_{ip}T_{jq} + T_{ip}L_{jq}) \cos\eta
	+ (\epsilon_{ipa}L_{jq}\hat q_a + \epsilon_{jqb}L_{ip}\hat q_b) \sin\eta \\
&\quad + \frac{1}{2} T_{ij} T_{pq} \cos 2\eta
	+ \frac{1}{2} (\epsilon_{jqa}T_{ip}\hat q_a 
	+ \epsilon_{ipb} T_{jq} \hat q_b) \sin 2\eta,
\ea
where we have introduced $L_{ij} = \hat q_i \hat q_j$ and $T_{ij} = \delta_{ij} - \hat q_i \hat q_j$, and used
\be
\label{eq:0517d}
\sum_{ab} \epsilon_{ipa} \epsilon_{jqb} \hat q_a \hat q_b = T_{ij}T_{pq} - T_{iq}T_{pj}.
\ee
We have also used the fact that, due to the structure of the tensors and vectors with which $L$ and $T$ are contracted, we are permitted to interchange the indices $i \leftrightarrow j$, $k \leftrightarrow l$, $p \leftrightarrow q$, and $r \leftrightarrow s$. (For this reason we use an arrow rather than an equality in \eqref{eq:0517a}.) Contracting with the appropriate tensors and averaging over the angle $\eta$, we obtain
\be
\label{eq:0517b}
\Lambda^m_{ij}(x \pm \tfrac{\hat q}{2})_p (x \pm \tfrac{\hat q}{2})_q
	\langle A_\eta^{ip} A_\eta^{jq} \rangle_\eta 
	= \Lambda^m_{ij}(x \pm \tfrac{\hat q}{2})_p (x \pm \tfrac{\hat q}{2})_q
	\bigg( L_{ij} L_{pq} + \frac{1}{2} T_{ij}T_{pq} \bigg),
\ee
\ba
\label{eq:0517c}
&\Lambda^m_{ij}\Lambda^n_{kl}
	(x + \tfrac{\hat q}{2})_p (x + \tfrac{\hat q}{2})_q 
	(x - \tfrac{\hat q}{2})_r (x - \tfrac{\hat q}{2})_s
	\langle A_\eta^{ip} A_\eta^{jq} A_\eta^{kr} A_\eta^{ls} \rangle_\eta \\
	&\quad \to \Lambda^m_{ij}\Lambda^n_{kl}
	(x + \tfrac{\hat q}{2})_p (x + \tfrac{\hat q}{2})_q 
	(x - \tfrac{\hat q}{2})_r (x - \tfrac{\hat q}{2})_s \bigg(
	L_{ij}L_{pq}L_{kl}L_{rs} + \frac{1}{2} L_{ij}L_{pq}T_{kl}T_{rs} \\
	&\quad\quad\quad\quad\quad + \frac{1}{2} L_{kl}L_{rs}T_{ij}T_{pq}
	+ 2 L_{ik}L_{pr}T_{jl}T_{qs}	+ \frac{1}{8} T_{ij}T_{pq}T_{kl}T_{rs}
	+ \frac{1}{4} T_{il}T_{ps}T_{jk}T_{qr} \bigg).
\ea
In obtaining this last expression, it was necessary to apply the angular rotation and averaging a second time in order to fully separate out the dependence on $\hat \bq$.

\subsection{Case of isotropic fermion dispersion}
Beginning with the fully isotropic case ($c_1 = c_2$ and $\alpha=0$), using the rotation of the integration variable described above, and taking traces with the appropriate matrices, the scaling functions appearing in \eqref{eq:1119a} are given by the following integrals over momentum:
\be
\label{eq:1105a}
f_1 \left( \frac{i\Omega}{\bq^2} \right) 
	= \frac{2}{(2\pi)^2} \int_0^\infty dr r^2 \left\{
	\frac{1}{4r^2} - \frac{1+4r^2}{4 \left[ \left( 2r^2 + \frac{1}{2} \right)^2
	-  \left( \frac{i\Omega}{\bq^2} \right)^2 \right]} \right\} = \frac{1}{64\pi} \left( \sqrt{1+2 \frac{i\Omega}{\bq^2}} + \sqrt{1- 2 \frac{i\Omega}{\bq^2}} \right),
\ee
\be
\label{eq:1105c}
f_3 \left( \frac{i\Omega}{\bq^2} \right) = \frac{1}{(2\pi)^2} \int_{-\infty}^\infty dz \int_0^\infty d\rho \; \rho
	\Bigg\{  \frac{\left( \rho^2 + z^2 + \frac{1}{4} \right)
	\left( z^4 - \frac{1}{2} z^2 - 3z^2 \rho^2 + \frac{1}{4} \rho^2 + \frac{3}{8}\rho^4 + \frac{1}{16} \right)}
	{\left[ \left( \rho^2 + z^2 + \frac{1}{4} \right)^2 - z^2 \right]
	\left[ \left( 2\rho^2 + 2z^2 + \frac{1}{2} \right)^2 - \left(\frac{i\Omega}{\bq^2}\right)^2 \right]} - \frac{z^4 - 3z^2 \rho^2 + \frac{3}{8} \rho^4}{4 (\rho^2 + z^2)^3}
	\Bigg\},
\ee
\be
\label{eq:1110aa}
f_4 \left( \frac{i\Omega}{\bq^2} \right) = \frac{1}{(2\pi)^2} \int_{-\infty}^\infty dz \int_0^\infty d\rho \; \rho^3
	\Bigg\{  \frac{\left( \rho^2 + z^2 + \frac{1}{4} \right)
	\left( 2 z^2 - \frac{1}{2} \rho^2 - \frac{1}{2} \right)}
	{\left[ \left( \rho^2 + z^2 + \frac{1}{4} \right)^2 - z^2 \right]
	\left[ \left( 2\rho^2 + 2z^2 + \frac{1}{2} \right)^2 - \left(\frac{i\Omega}{\bq^2}\right)^2 \right]} - \frac{2 z^2 - \frac{1}{2} \rho^2}{4 (\rho^2 + z^2)^3}
	\Bigg\},
\ee
\be
\label{eq:1110a}
f_5 \left( \frac{i\Omega}{\bq^2} \right) = \frac{1}{(2\pi)^2} \int_{-\infty}^\infty dz \int_0^\infty d\rho \; \rho^5
	\Bigg\{  \frac{\left( \rho^2 + z^2 + \frac{1}{4} \right)}
	{4\left[ \left( \rho^2 + z^2 + \frac{1}{4} \right)^2 - z^2 \right]
	\left[ \left( 2\rho^2 + 2z^2 + \frac{1}{2} \right)^2 - \left(\frac{i\Omega}{\bq^2}\right)^2 \right]}  - \frac{1}{16 (\rho^2 + z^2)^3}
	\Bigg\}.
\ee
The integration coordinates are related to the original momenta $\bk$ and $\bq$ as $z = \bk\cdot\bq/\bq^2$ and $\rho^2 = \bk^2/\bq^2 - z^2$. In \eqref{eq:1105c}--\eqref{eq:1110a} it is possible to perform the integrals over $z$ numerically, then it is straightforward to perform the remaining integrals numerically, as they are UV convergent due to the subtraction of $\hat\Pi_\phi (0,0)$ in \eqref{eq:1119a}.

In the case of isotropic fermion dispersion ($c_1 = c_2$) but arbitrary Yukawa parameter $\alpha$, the scaling functions determining the polarization tensor in \eqref{eq:0514a} can be expressed as the following linear combinations of $f_i(\tfrac{i\Omega}{\bq^2})$ in \eqref{eq:1105a}--\eqref{eq:1110a}:
\ba
\label{eq:0515c}
h_1 (\tfrac{i\Omega}{\bq^2}, \alpha) =& 
	\frac{5}{64} ( 178 f_1 + 22 f_3 + 22 f_4 + 267 f_5)
	-\frac{3}{64} (190 f_1 + 58 f_3 + 58 f_4 + 285 f_5) \cos(2\alpha) \\
	&+ \frac{1}{8} (82 f_1 - 2 f_3 - 2 f_4 + 123 f_5) \sin(2\alpha),
\ea
\be
\label{eq:0515d}
h_2 (\tfrac{i\Omega}{\bq^2}, \alpha) = 
	\frac{9}{8} (26 f_3 + 17 f_4) \left[ \cos (2\alpha) - 1 \right]
	- 9(f_3 + f_4) \sin(2\alpha) ,
\ee
\be
\label{eq:0515e}
h_3 (\tfrac{i\Omega}{\bq^2}, \alpha) =
	\frac{81}{4} f_3 \left[ 1-\cos(2\alpha) \right],
\ee
\be
\label{eq:0515f}
h_4 (\tfrac{i\Omega}{\bq^2}, \alpha) =
	\frac{3}{64} (404 f_3 + 743 f_4)
	- \frac{9}{64} (92 f_3 + 173 f_4) \cos(2\alpha)
	+ \frac{3}{8} (28 f_3 + 67 f_4) \sin(2\alpha),
\ee
\be
\label{eq:0515g}
h_5 (\tfrac{i\Omega}{\bq^2}, \alpha) =
	\frac{9}{8} f_3 \left[ 1 - \cos(2\alpha) \right] 
	+ \frac{9}{2} f_3 \sin(2\alpha).
\ee

\subsection{Fully anisotropic case}
In the more general anisotropic case where $c_1 \neq c_2$, we introduce the anisotropy parameter $\delta$ for the fermion dispersion by letting $c_{1,2} = 1 \pm \delta$. The integrands in \eqref{eq:0220g}--\eqref{eq:0220j} can then be expanded perturbatively to leading order in $\delta$. The fermion dispersion becomes
\be
\label{eq:1125a}
\varepsilon_{\bx \pm \hat\bk/2} = \left(\bx \pm \tfrac{\hat\bk}{2} \right)^2 
	+ \delta \frac{\sum_{m=1}^5 \xi_m d_m^2(\bx \pm \frac{\hat\bk}{2})}{(\bx \pm \frac{\hat\bk}{2})^2}
	+ O(\delta^2) ,
\ee 
where we have introduced $\xi_m = (1,1,1,-1,-1)_m$. As in the isotropic case, we can separate out the dependence on $\hat\bq$ in \eqref{eq:0220g}--\eqref{eq:0220j} by repeatedly applying the rotation \eqref{eq:1029d} and averaging over the angle $\eta$. (In some cases products of more than four rotation matrices appear, leading to a much longer and more tedious calculation in which the transformation \eqref{eq:1029d} must be performed up to four times.) This procedure results in the following expressions:
\be
\label{eq:1221e}
\Phi_1 \left(\tfrac{i\Omega}{\bq^2}, \hat\bq \right) =
	\Phi_{1,1} \left(\tfrac{i\Omega}{\bq^2}\right)
	+ \delta \left[ \Phi_{1,2} \left(\tfrac{i\Omega}{\bq^2} \right)
	+ \Phi_{1,3} \left(\tfrac{i\Omega}{\bq^2} \right) 
	\sum_a \xi_a (\hat\bq \cdot \Lambda^a \cdot \hat\bq)^2 \right]
	+ O(\delta^2),
\ee
\ba
\label{eq:1222c}
\Phi_2^{(n)} \left( \tfrac{i\Omega}{\bq^2}, \hat\bq \right) &= 
	\Phi_{2,1} \left( \tfrac{i\Omega}{\bq^2} \right) 
	\hat\bq\cdot\Lambda^n\cdot\hat\bq
	+ \delta \bigg[ \Phi_{2,2} \left( \tfrac{i\Omega}{\bq^2} \right)
	\hat\bq\cdot\Lambda^n\cdot\hat\bq
	+ \Phi_{2,3} \left( \tfrac{i\Omega}{\bq^2} \right)
	\hat\bq\cdot\Lambda^n\cdot\hat\bq
	\sum_a \xi_a (\hat\bq\cdot\Lambda^a\cdot\hat\bq)^2 \\
&\quad + \Phi_{2,4} \left( \tfrac{i\Omega}{\bq^2} \right)
	\sum_a \xi_a (\hat\bq\cdot\Lambda^a\cdot\hat\bq)
	(\hat\bq\cdot\Lambda^a\Lambda^n\cdot\hat\bq)
	+ \Phi_{2,5} \left( \tfrac{i\Omega}{\bq^2} \right)
	\sum_a \xi_a 
	\hat\bq\cdot\Lambda^a\Lambda^n\Lambda^a\cdot\hat\bq
	\bigg] + O(\delta^2), 
\ea
\ba
\label{eq:1224k}
\Phi_4^{(m,n)}& \left( \tfrac{i\Omega}{\bq^2} , \hat\bq \right) =
	\Phi_{4,1} \left( \tfrac{i\Omega}{\bq^2} \right) 
	(\hat\bq\cdot\Lambda^m\cdot\hat\bq)(\hat\bq\cdot\Lambda^n\cdot\hat\bq)
	+ \Phi_{4,2} \left( \tfrac{i\Omega}{\bq^2} \right)
	\hat\bq\cdot\Lambda^m\Lambda^n\cdot\hat\bq
	+ \Phi_{4,3} \left( \tfrac{i\Omega}{\bq^2} \right)
	\mathrm{Tr}[\Lambda^m \Lambda^n ] \\
& + \delta \bigg\{ \left[ \Phi_{4,4}^{(0)} \left( \tfrac{i\Omega}{\bq^2} \right)
	+ (\xi_m + \xi_n) \Phi_{4,4}^{(1)} \left( \tfrac{i\Omega}{\bq^2} \right)
	\right]
	 (\hat\bq\cdot\Lambda^m\cdot\hat\bq)
	(\hat\bq\cdot\Lambda^n\cdot\hat\bq) \\
&\quad\quad + \left[ \Phi_{4,5}^{(0)} \left( \tfrac{i\Omega}{\bq^2} \right)
	+ (\xi_m + \xi_n) \Phi_{4,5}^{(1)} \left( \tfrac{i\Omega}{\bq^2} \right)
	\right]
	\hat\bq\cdot\Lambda^m\Lambda^n\cdot\hat\bq \\
&\quad\quad  + \left[ \Phi_{4,6}^{(0)} \left( \tfrac{i\Omega}{\bq^2} \right)
	+ (\xi_m + \xi_n) \Phi_{4,6}^{(1)} \left( \tfrac{i\Omega}{\bq^2} \right)
	\right] \mathrm{Tr} [\Lambda^m \Lambda^n] 
	+ \Phi_{4,7} \left( \tfrac{i\Omega}{\bq^2} \right)
	 (\hat\bq\cdot\Lambda^m\Lambda^n\cdot\hat\bq) 
	\sum_a \xi_a (\hat\bq\cdot\Lambda^a\cdot\hat\bq)^2 \\
&\quad\quad + \Phi_{4,8} \left( \tfrac{i\Omega}{\bq^2} \right) 
	(\hat\bq\cdot\Lambda^n\cdot\hat\bq)
	\sum_a \xi_a (\hat\bq\cdot\Lambda^a\cdot\hat\bq)
	(\hat\bq\cdot\Lambda^a\Lambda^m\cdot\hat\bq) \\
&\quad\quad + \Phi_{4,9} \left( \tfrac{i\Omega}{\bq^2} \right)
	(\hat\bq\cdot\Lambda^m\cdot\hat\bq)
	\sum_a \xi_a (\hat\bq\cdot\Lambda^a\cdot\hat\bq)
	(\hat\bq\cdot\Lambda^a\Lambda^n\cdot\hat\bq) \\
&\quad\quad + \Phi_{4,10} \left( \tfrac{i\Omega}{\bq^2} \right)
	\sum_a \xi_a (\hat\bq\cdot\Lambda^a\Lambda^m\cdot\hat\bq)
	(\hat\bq\cdot\Lambda^a\Lambda^n\cdot\hat\bq)
	+ \Phi_{4,11} \left( \tfrac{i\Omega}{\bq^2} \right)
	\sum_a \xi_a (\hat\bq\cdot\Lambda^a\cdot\hat\bq)
	(\hat\bq\cdot\Lambda^m\Lambda^a\Lambda^n\cdot\hat\bq) \\
&\quad\quad + \Phi_{4,12} \left( \tfrac{i\Omega}{\bq^2} \right)
	\sum_a \xi_a (\hat\bq\cdot\Lambda^a\cdot\hat\bq)
	(\hat\bq\cdot\Lambda^m\Lambda^n\Lambda^a\cdot\hat\bq)
	+ \Phi_{4,13} \left( \tfrac{i\Omega}{\bq^2} \right)
	\sum_a \xi_a (\hat\bq\cdot\Lambda^a\cdot\hat\bq)
	(\hat\bq\cdot\Lambda^n\Lambda^m\Lambda^a\cdot\hat\bq) \\
&\quad\quad + \Phi_{4,14} \left( \tfrac{i\Omega}{\bq^2} \right)
	(\hat\bq\cdot\Lambda^n\cdot\hat\bq) \sum_a \xi_a 
	(\hat\bq\cdot\Lambda^a\Lambda^m\Lambda^a\cdot\hat\bq)
	+ \Phi_{4,15} \left( \tfrac{i\Omega}{\bq^2} \right)
	(\hat\bq\cdot\Lambda^m\cdot\hat\bq) \sum_a \xi_a 
	(\hat\bq\cdot\Lambda^a\Lambda^n\Lambda^a\cdot\hat\bq) \\
&\quad\quad + \Phi_{4,16} \left( \tfrac{i\Omega}{\bq^2} \right)
	\sum_a \xi_a \hat\bq\cdot\Lambda^a\Lambda^m\Lambda^n\Lambda^a\cdot\hat\bq
	+ \Phi_{4,17} \left( \tfrac{i\Omega}{\bq^2} \right)
	\sum_a \xi_a \hat\bq\cdot\Lambda^m\Lambda^a\Lambda^n\Lambda^a\cdot\hat\bq \\
&\quad\quad + \Phi_{4,18} \left( \tfrac{i\Omega}{\bq^2} \right)
	\sum_a \xi_a \hat\bq\cdot\Lambda^n\Lambda^a\Lambda^m\Lambda^a\cdot\hat\bq  
	+ \Phi_{4,19} \left( \tfrac{i\Omega}{\bq^2} \right) 	
	 (\hat\bq\cdot\Lambda^m\cdot\hat\bq)
	(\hat\bq\cdot\Lambda^n\cdot\hat\bq) \sum_a \xi_a
	(\hat\bq\cdot\Lambda^a\cdot\hat\bq)^2 \\
&\quad\quad	+ \Phi_{4,20} \left( \tfrac{i\Omega}{\bq^2} \right) 
	\mathrm{Tr} [\Lambda^m \Lambda^n]
	\sum_a \xi_a (\hat\bq\cdot\Lambda^a\cdot\hat\bq)^2
	+ \Phi_{4,21} \left( \tfrac{i\Omega}{\bq^2} \right) 
	\sum_a \xi_a \mathrm{Tr} [ \Lambda^m \Lambda^n \Lambda^a ] 
	\hat\bq\cdot\Lambda^a\cdot\hat\bq
	\bigg\}	+ O(\delta^2).
\ea
The scaling functions appearing in \eqref{eq:1221e}--\eqref{eq:1224k} are similar in form to those given in the isotropic case, though we shall not list them all here explicitly for the sake of brevity.

Once the scaling functions are known, taking the traces in \eqref{eq:0928h} gives the following expressions for the components of the polarization tensor:
\ba
\label{eq:1130c}
\Pi_\phi^{xx} (i\Omega, \bq) - \Pi_\phi^{xx} (0,0) 
	= \frac{u^2}{32} & |\bq| \bigg[ 
	(328 \sin (2 \alpha )-285 \cos (2 \alpha )+445 ) 
	\Phi_1 (\tfrac{i\Omega}{\bq^2}, \hat\bq) \\
&+ (120 \sin (2 \alpha )+21 \cos (2 \alpha )+75)
	\left( \Phi_4^{(2,2)}(\tfrac{i\Omega}{\bq^2}, \hat\bq)
	+ \Phi_4^{(3,3)}(\tfrac{i\Omega}{\bq^2}, \hat\bq) 
	-\Phi_4^{(1,1)}(\tfrac{i\Omega}{\bq^2}, \hat\bq) \right) \\
&+ (216 \sin (2 \alpha )-219 \cos (2 \alpha )+315) 
	\Phi_4^{(4,4)} (\tfrac{i\Omega}{\bq^2}, \hat\bq) \\
&+ \sqrt{3} (-224 \sin (2 \alpha )+132 \cos (2 \alpha )-260)
	\Phi_4^{(4,5)} (\tfrac{i\Omega}{\bq^2}, \hat\bq) \\
&+ (-8 \sin (2 \alpha )-87 \cos (2 \alpha )+55 ) 
	\Phi_4^{(5,5)} (\tfrac{i\Omega}{\bq^2}, \hat\bq)
	\bigg],
\ea
\ba
\label{eq:1130d}
\Pi_\phi^{xy} (i\Omega, \bq) - \Pi_\phi^{xy} (0,0) 
	&= \frac{u^2}{16} |\bq| \bigg[3 (72 \sin (2 \alpha )-73 \cos (2 \alpha )+105) 
	\Phi_4^{(1,2)} (\tfrac{i\Omega}{\bq^2}, \hat\bq) \\
&\quad\quad\quad +2 \sqrt{3} (-80 \sin (2 \alpha )+75 \cos (2 \alpha )-107) 
	\Phi_4^{(3,5)} (\tfrac{i\Omega}{\bq^2}, \hat\bq) 
	\bigg],
\ea
\ba
\label{eq:1130e}
\Pi_\phi^{xz} (i\Omega, \bq) - \Pi_\phi^{xz} (0,0) 
	= \frac{u^2}{16} |\bq| \bigg[ &
	(216 \sin (2 \alpha )-219 \cos (2 \alpha )+315)
	\Phi_4^{(1,3)} (\tfrac{i\Omega}{\bq^2}, \hat\bq) \\
& + (240 \sin (2 \alpha )-225 \cos (2 \alpha )+321)
	\Phi_4^{(2,4)} (\tfrac{i\Omega}{\bq^2}, \hat\bq) \\
& + \sqrt{3} (80 \sin (2 \alpha )-75 \cos (2 \alpha )+107)
	\Phi_4^{(2,5)} (\tfrac{i\Omega}{\bq^2}, \hat\bq)
	\bigg],
\ea
\ba
\label{eq:1130f}
\Pi_\phi^{yy} (i\Omega, \bq) - \Pi_\phi^{yy} (0,0) 
	= \frac{u^2}{32} & |\bq| \bigg[ 
	(328 \sin (2 \alpha )-285 \cos (2 \alpha )+445)
	\Phi_1 (\tfrac{i\Omega}{\bq^2}, \hat\bq) \\
& + 3 (40 \sin (2 \alpha )+7 \cos (2 \alpha )+25)
	\left( \Phi_4^{(1,1)} (\tfrac{i\Omega}{\bq^2}, \hat\bq)
	- \Phi_4^{(2,2)} (\tfrac{i\Omega}{\bq^2}, \hat\bq)
	+ \Phi_4^{(3,3)} (\tfrac{i\Omega}{\bq^2}, \hat\bq) \right) \\
& + (216 \sin (2 \alpha )-219 \cos (2 \alpha )+315)
	\Phi_4^{(4,4)} (\tfrac{i\Omega}{\bq^2}, \hat\bq) \\
& + 4 \sqrt{3} (56 \sin (2 \alpha )-33 \cos (2 \alpha )+65)
	\Phi_4^{(4,5)} (\tfrac{i\Omega}{\bq^2}, \hat\bq) \\
& + (-8 \sin (2 \alpha )-87 \cos (2 \alpha )+55)
	\Phi_4^{(5,5)} (\tfrac{i\Omega}{\bq^2}, \hat\bq)
	\bigg],
\ea
\ba
\label{eq:1130g}
\Pi_\phi^{yz} (i\Omega, \bq) - \Pi_\phi^{yz} (0,0) 
	= \frac{u^2}{16} |\bq| \bigg[ &
	(-240 \sin (2 \alpha )+225 \cos (2 \alpha )-321)
	\Phi_4^{(1,4)} (\tfrac{i\Omega}{\bq^2}, \hat\bq) \\
& + \sqrt{3} (80 \sin (2 \alpha )-75 \cos (2 \alpha )+107)
	\Phi_4^{(1,5)} (\tfrac{i\Omega}{\bq^2}, \hat\bq) \\
& + (216 \sin (2 \alpha )-219 \cos (2 \alpha )+315)
	\Phi_4^{(2,3)} (\tfrac{i\Omega}{\bq^2}, \hat\bq)
	\bigg],
\ea
\ba
\label{eq:1130h}
\Pi_\phi^{zz} (i\Omega, \bq) - \Pi_\phi^{zz} (0,0) 
	= \frac{u^2}{32} |\bq| \bigg[ &
	(328 \sin (2 \alpha )-285 \cos (2 \alpha )+445)
	\left( \Phi_1 (\tfrac{i\Omega}{\bq^2}, \hat\bq)
	+ \Phi_4^{(5,5)} (\tfrac{i\Omega}{\bq^2}, \hat\bq) \right) \\
& + (120 \sin (2 \alpha )+21 \cos (2 \alpha )+75)
	\Big( \Phi_4^{(1,1)} (\tfrac{i\Omega}{\bq^2}, \hat\bq)
	+ \Phi_4^{(2,2)} (\tfrac{i\Omega}{\bq^2}, \hat\bq) \\
&\quad\quad\quad\quad\quad\quad\quad\quad\quad\quad\quad\quad
	- \Phi_4^{(3,3)} (\tfrac{i\Omega}{\bq^2}, \hat\bq)
	- \Phi_4^{(4,4)} (\tfrac{i\Omega}{\bq^2}, \hat\bq) \Big) 
	\bigg].
\ea
The remaining off-diagonal components follow from $\Pi_\phi^{ij} = \Pi_\phi^{ji}$.

\end{widetext}

\end{appendix}

\bibliographystyle{apsrev}
\bibliography{refs}

\end{document}